\def\year{2021}\relax
\documentclass[letterpaper]{article} % DO NOT CHANGE THIS
\usepackage{aaai21}  % DO NOT CHANGE THIS
\usepackage{times}  % DO NOT CHANGE THIS
\usepackage{helvet} % DO NOT CHANGE THIS
\usepackage{courier}  % DO NOT CHANGE THIS
\usepackage[hyphens]{url}  % DO NOT CHANGE THIS
\usepackage{graphicx} % DO NOT CHANGE THIS
\usepackage{appendix}
\usepackage{natbib}  % DO NOT CHANGE THIS AND DO NOT ADD ANY OPTIONS TO IT
\usepackage{caption} % DO NOT CHANGE THIS AND DO NOT ADD ANY OPTIONS TO IT
\usepackage{amsfonts}
\usepackage{amssymb}
\usepackage{amsmath}
\usepackage{algpseudocode,algorithm,algorithmicx}
\usepackage{array}
\usepackage{booktabs}
\usepackage{multirow}
\usepackage{color}
\usepackage{subfigure}
\DeclareMathOperator*{\argmax}{arg\,max}

\usepackage{textcomp,booktabs}
\usepackage{colortbl}
\usepackage[switch]{lineno}  

\title{Composite Adversarial Attacks}
\author {
        Xiaofeng Mao,\textsuperscript{\rm 1}
        Yuefeng Chen, \textsuperscript{\rm 1}
        Shuhui Wang, \textsuperscript{\rm 2}\thanks{Shuhui Wang is the corresponding author.}
        Hang Su, \textsuperscript{\rm 3}
        Yuan He, \textsuperscript{\rm 1}
        Hui Xue \textsuperscript{\rm 1}
        \\
}
\affiliations {
    \textsuperscript{\rm 1} Alibaba Group,
    \textsuperscript{\rm 2} Inst. of Comput. Tech., CAS
    \\
    \textsuperscript{\rm 3} Tsinghua University 
    \\
    
    \{mxf164419, yuefeng.chenyf\}@alibaba-inc.com, wangshuhui@ict.ac.cn, suhangss@mail.tsinghua.edu.cn
}

\begin{document}
\maketitle

\begin{abstract}
Adversarial attack is a technique for deceiving Machine Learning (ML) models, which provides a way to evaluate the adversarial robustness. In practice, attack algorithms are artificially selected and tuned by human experts to break a ML system. However, manual selection of attackers tends to be sub-optimal, leading to a mistakenly assessment of model security. In this paper, a new procedure called Composite Adversarial Attack (CAA) is proposed for automatically searching the best combination of attack algorithms and their hyper-parameters from a candidate pool of \textbf{32 base attackers}. We design a search space where attack policy is represented as an attacking sequence, {\it i.e.}, the output of the previous attacker is used as the initialization input for successors. Multi-objective NSGA-II genetic algorithm is adopted for finding the strongest attack policy with minimum complexity. The experimental result shows CAA beats 10 top attackers on 11 diverse defenses with less elapsed time (\textbf{6 $\times$ faster than AutoAttack}), and achieves the new state-of-the-art on $l_{\infty}$, $l_{2}$ and unrestricted adversarial attacks.
% AutoAttack achieves the state-of-the-art on the trade-off on attack performance and cost on 10 defenses. Besides, it breaks through all existing defense models under unrestricted setting with nearly 100\% success rate.
\end{abstract}

\section{Introduction} 
DNNs are vulnerable towards adversarial attacks, which aim to fool a well trained model by producing imperceptibly perturbed examples. This serious security implication quickly attracted a lot of attention from the machine learning community. With in-depth study of adversarial examples, a lot of attack algorithms are proposed to validate the adversarial robustness. Meanwhile, several open source toolboxes, such as Cleverhans\footnotemark~\cite{papernot2016cleverhans}, FoolBox\footnotemark~\cite{rauber2017foolbox} or AdverTorch\footnotemark~\cite{ding2019advertorch}, are developed and integrating most existing attack algorithms. All of them provided user friendly interface for attacking a model conveniently and quickly.

\begin{figure}[!htb]
\centering
\includegraphics[width=8cm]{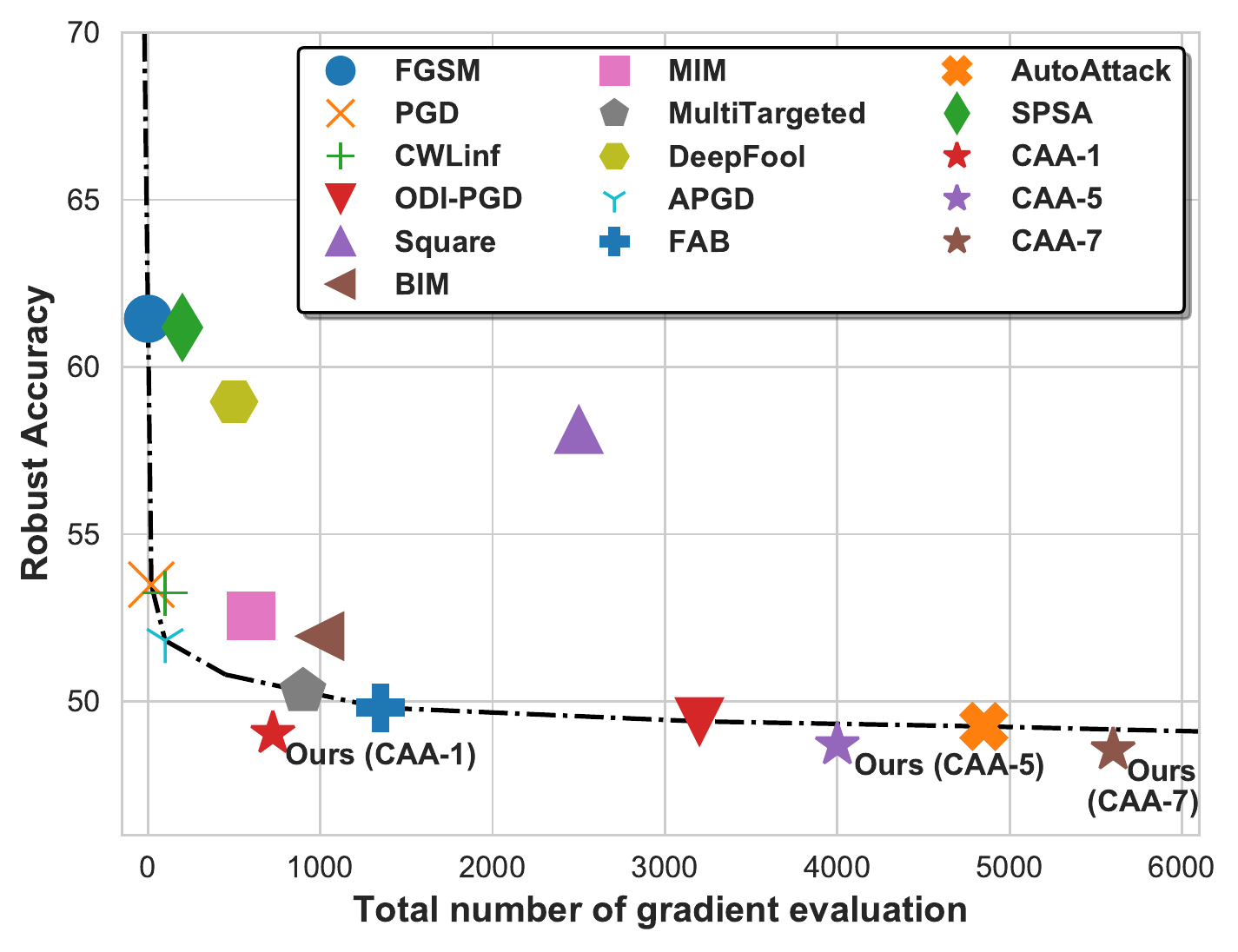}

\caption{Comparison between CAA and state-of-the-art attackers on CIFAR-10 adversarial training model. CAA-$n$ represents the CAA attack with $n$ restarts. Our method achieves the best attack performance only with a small number of gradient evaluation. }
\label{fig:intro1}
\end{figure}

However, even if well-designed toolboxes are developed, attacking a model still needs a lot of user experience or manual tuning of hyper-parameters of the attacks, especially when we cannot know the defense mechanism of the target model. This user-dependent characteristic also makes it hard for instrumentalization of adversarial attacks. On the other hand, manually selecting attackers is somewhat tendentious and sub-optimal. It may arouse a mistakenly assessment of model security, {\it e.g.}, a well-known false sense of security is gradient obfuscation, which leads to illusory defense for gradient-based attacks. 

\begin{figure}[!htb]
\centering
\includegraphics[width=8.3cm]{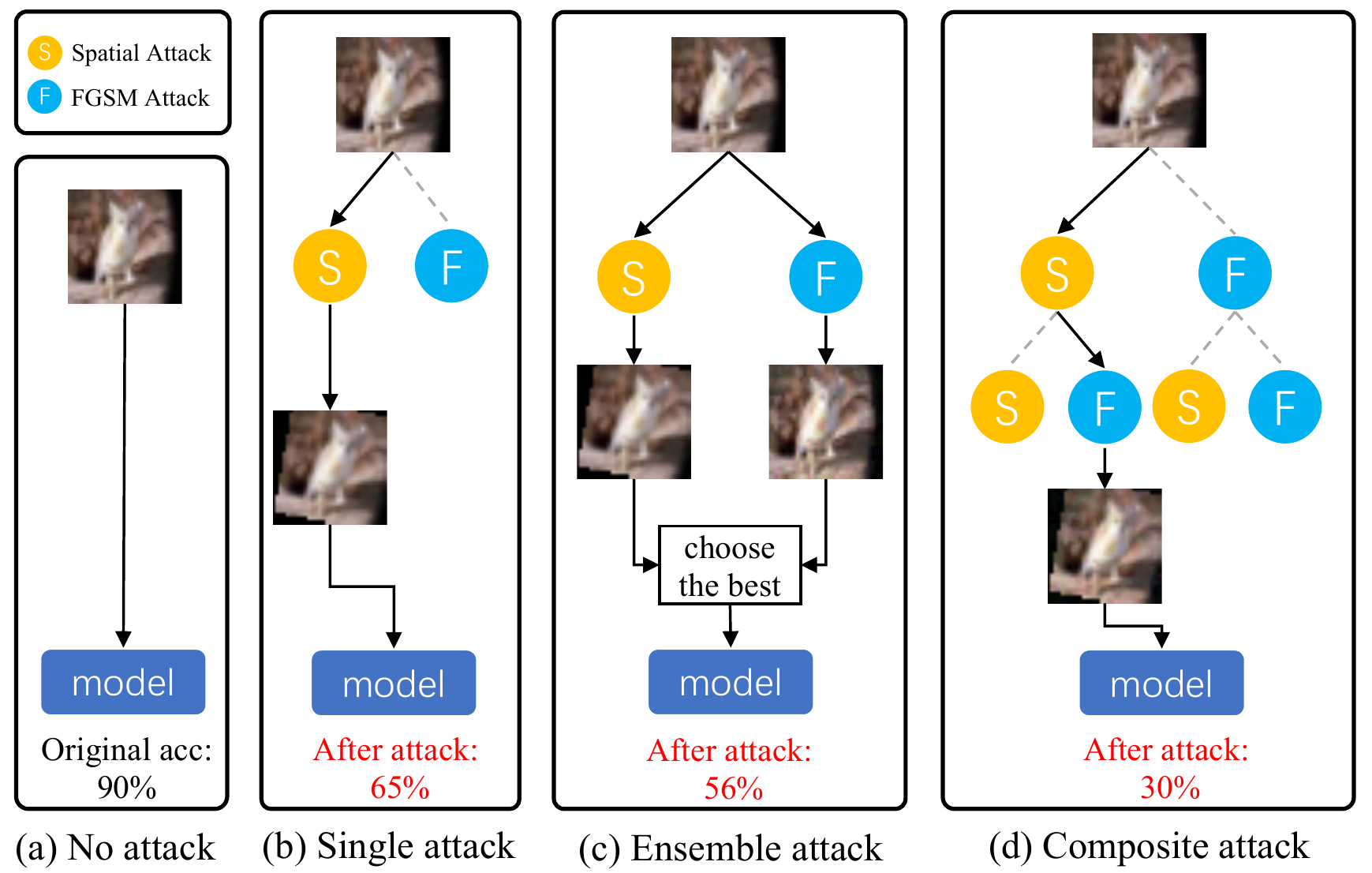}

\caption{Illustration of single attack, ensemble attack and our composite attack. $S$ and $F$ denote Spatial and FGSM attack, respectively.}
\label{fig:intro2}
\end{figure}

In order to perform more comprehensive and stronger attacks, we first propose to automate the attack process by searching an effective attack policy from a collection of attack algorithms. 
% (e.g., FGSM~\cite{goodfellow2014explaining}, PGD~\cite{madry2017towards} or C\&W~\cite{carlini2017towards}). 
We name this procedure as Composite Adversarial Attacks (CAA). To demonstrate the key idea of CAA, an example is presented in Fig.~\ref{fig:intro2}. Suppose that there are two candidate ways of attacks, {\it i.e.}, Spatial attack~\cite{engstrom2019exploring} and FGSM attack~\cite{goodfellow2014explaining}, the goal is choosing one or more of them to compose a stronger attack policy. In Fig.~\ref{fig:intro2} (b), the simplest way is selecting the best single attack as the final policy. 
\footnotetext[1]{https://github.com/tensorflow/cleverhans}
\footnotetext[2]{https://github.com/bethgelab/foolbox}
\footnotetext[3]{https://github.com/BorealisAI/advertorch}
However, single attacker is always not strong and generalized enough as shown in previous works~\cite{tramer2019adversarial}. A more potential solution~\cite{croce2020reliable} is to find multiple attackers, and then ensemble them by choosing the best output constantly that can successfully fool the model (Fig.~\ref{fig:intro2} (c)). Although higher attack success rate can be obtained, the ensemble attack only provides the output level aggregation, without considering the complementarity among different attacking mechanisms.

% The simplest way is selecting the best performed attack, as shown in Figure~\ref{fig:intro2} (a). 
% Figure~\ref{fig:intro2} presents three ways of searching attackers from the collection. The most common and simple practice is shown in Figure~\ref{fig:intro2} (a), in which a best performed attack algorithm (Spatial) is selected to attack the model. However, single attacker is always not strong and generalized enough claimed in previous works~\cite{tramer2019adversarial,liu2018caad}. Therefore, another alternative way is finding a subset of attackers, then ensemble them.  by considering the worst case over all of attacks (Figure~\ref{fig:intro2} (b)).
% A lot of previous works~\cite{croce2020reliable} have used such an ensemble way, which can be regarded as a parallel processing and merging of attackers.
% Hypothesis 1 (H1): Local adversarial examples are better
% starting points for optimization attacks than original seeds.
% Liu et al. observe that for the same classification tasks, different models tend to have similar decision boundaries [29].
% Therefore, we hypothesize that, although candidate adversarial examples generated on local models may not fully transfer
% to the target model, these candidates are still closer to the
% targeted region than the original seed and hence, make better
% starting points for optimization attacks.
% We can regarded the ensemble attack as a parallel processing and merging of attackers. 

%we use a different but advanced way to form the attack policy.

In our composite adversarial attack, We define an attack policy as the serial connection of attackers, in which the output of previous attacker is used as the initialization input for successors. In Fig.~\ref{fig:intro2} (d), four possible permutations can be generated by two attackers. By using a search algorithm to find the best permutation, we show that 
an FGSM attack following Spatial attack can achieve 26\% higher error rate than ensemble of them. The advantage of our policy lies in two aspects: 1) By introducing identity attack ({\it i.e.}, no attack), our CAA can represent any single attack. Ensemble attack can also be represented by policy ensemble of CAA. Therefore, CAA is the more generalized formulation. 2) A strong attack can be produced via progressive steps. Previous works~\cite{suya2020hybrid} have found that some starting points close to the decision boundary are better than the original seeds for optimizing the attacks. Similarly in CAA, we use preceding attackers to create an example far enough from the original seed and close enough to the boundary, such that subsequent attacks are easier to find a stronger adversarial example. 

%To further validate the effectiveness of CAA, we make an empirical comparison of these several ways for the attack policy design in the experiment.

% expresses  several  choices  and orders of  attackers
% that serialize the attackers. In Figure
% Relatively, there is a third way that serialize the attackers instead of parallelization. We called it composite attack, in which the output of the previous attacker is used as the initialization input for successors, forming an attack sequence. Composite attack can gradually accumulates different types of perturbations, causing the target model gets the worst accuracy (30.30\%). Also a carefully comparison of these several ways for the attack policy design is provided in Sec~\ref{sec:exp1}. Overall, we adopt the consecutive sequence of multiple attackers as a way to form the attack policy in this work.

Specifically, CAA is implemented with a search space containing several choices and orders of attack operations. For each attack operation, there are two hyper-parameters, {\it i.e.}, magnitude $\epsilon$ and iteration steps $t$. We adopt NSGA-II genetic algorithm~\cite{deb2002fast} to find the best attack policy which can break through the target model with highest success rate but have the minimal complexity. Extensive experiments show that CAA achieves excellent improvements in two use cases: 1) CAA can be applied directly on the target model of interest to find the best attack policy (CAA$_{dic}$) and 2) learned policies can keep high success rate transferred to attack multiple model architectures, under different tasks (CAA$_{sub}$). We evaluate CAA$_{dic}$ and CAA$_{sub}$ on 11 recent proposed defenses on $l_{\infty}$, $l_{2}$ and unrestricted setting. The result shows our composite adversarial attack achieves the new state-of-the-art in white-box scenario, with a signiﬁcant reduction of attack time cost.

\section{Preliminaries and Related work}
\subsection{Adversarial Attack} 
\subsubsection{Definition and Notation}
Let $\mathcal{F}: x\in [0,1]^{D} \rightarrow z\in \mathbb{R}^{K}$ be a $K$-class image classifier, where $x$ is an input image in the $D$-dimensional image space, $z$ represents the logits. Suppose $\mathcal{F}$ is well performed and correctly classify $x$ as its ground truth label $y$. The purpose of adversarial attacks is to find an adversarial example $x_{adv}$ which is close to the original $x$ under a certain distance metric but causes mis-classification of the model:  $\mathcal{F}(x_{adv}) \neq y $. 

\subsubsection{Regular adversarial examples}
Regular adversarial examples are with limited magnitude of perturbations, which is always achieved by bounding the perturbations within the $\epsilon$-radius $l_{p}$-ball around the input $x$. It can be formed by $\mathcal{F}(x_{adv}) \neq y \ s.t. \ \left \| x_{adv}-x \right \|_{p}\leq \lambda$. Fast Gradient Sign Method (FGSM) is a classic $l_{\infty}$ adversarial attack approach performing single step update on the original sample $x$ along the direction of the gradient of loss function. There are many improved versions of FGSM using momentum based multi-steps optimization~\cite{dong2018boosting}, or random initialization of perturbations~\cite{madry2017towards}. 
$l_{2}$ based attacks such as DDNL2~\cite{rony2019decoupling} and C\&W~\cite{carlini2017towards} find $x_{adv}$ which has the minimal $l_{2}$ distance to the its original examples. 
$l_{1}$ based attackers guarantee the sparsity of the perturbation, such as EAD~\cite{chen2017ead}.
However, $l_{1}$ attacks are not commonly used in practical attack setting. Therefore, we have not implemented CAA under $l_{1}$ constraint in this paper.

% optimized There are multiple formalities of adversarial example based on different values of $p$, such as $l_{\infty}$-based (), $$

% Then an attack algorithm find optimal $x_{adv}$ to maximize the classification error under this constrain: $\argmax \sum_ $

\subsubsection{Unrestricted adversarial examples}
Unrestricted adversarial example is a new type of adversarial example which is not restricted to small norm bounded perturbations. In this case, the attacker might change an input significantly without changing the semantics. \cite{brown2018unrestricted} first introduces the concept of unrestricted adversarial examples and raises a two-player unrestricted attack\&defense contest. Recently, there are lots of works aiming to construct such a stronger unrestricted attack using generative models~\cite{song2018constructing} or spatial transforms~\cite{engstrom2019exploring}. In this paper, we also implement unrestricted CAA with the largest search space (overall 19 attacks). We found that even applying very simple base attackers to form the search space, the policy searched by our CAA still yields surprising attack ability at unrestricted setting.

\subsection{Automated Machine Learning}
Our approach is inspired by recent advances in AutoML and its sub-directions such as Neural Architecture Search (NAS) and Hyper-parameter Optimization (HPO). In AutoML, search algorithms are used for choice of algorithm, feature pre-processing steps and hyper-parameters automatically. Another similar direction is AutoAugment~\cite{cubuk2018autoaugment}, which automatically searches for improved data augmentation policies. These automation technologies not only make people get rid of the tedious process of algorithm fine-tuning, but also greatly improve the effect and efficiency of the learning system. In this work, we adopts some search technologies in AutoML, and demonstrates that searching better algorithms and parameters also helps for the adversarial attacks.

\section{Composite Adversarial Attacks}
\subsection{Problem formulation}
Assume that we have an annotated dataset $\{X,Y\}$ and a collection of attack algorithms with some unknown hyper-parameters. In this paper, each attack algorithm is regard as an operation $\mathcal{A}: x\in[0,1]^{D}\rightarrow x_{adv}\in[0,1]^{D}$, which transforms the input $x$ to adversarial one $x_{adv}$ on the image space. $\mathcal{A}$ has various choices under different attack settings. For example, in white-box adversarial attack, $\mathcal{A}$ directly optimizes a perturbation $\delta $ within the $\epsilon$-radius ball around the input $x$, for maximizing the classification error:

\begin{equation}
    \mathcal{A}(x, \mathcal{F};\epsilon)= \argmax_{x+\delta}L(\mathcal{F}(x+\delta),y) \ \ s.t. \left \| \delta \right \|_{p}\leq \epsilon,
\end{equation}
where $L$ typically refers to the cross entropy loss, $\left \| \cdot  \right \|_{p}$ presents the $l_{p}$-norm and $\epsilon$ is the bound of the $l_{p}$-norm. $\epsilon$ can be viewed as a hyper-parameter of $\mathcal{A}$. Besides, there are many other attack settings, {\it e.g.}, black-box attack~\cite{uesato2018adversarial, andriushchenko2019square}, unrestricted adversarial attack~\cite{brown2018unrestricted}. We represent them as the attack operation $\mathcal{A}$ in a unified way.

Suppose we have a set of base attack operations, presented as $\mathbb{A}=\{\mathcal{A}_{1}, \mathcal{A}_{2}...\mathcal{A}_{k}\}$, where $k$ is the total number of attack operations. The goal of composite adversarial attack is to automate the adversarial attack process by searching for the best composition of attack operations in $\mathbb{A}$ and hyper-parameters of each operation, to achieve more general and powerful attacks. In this work, we only consider two most common hyper-parameters in attack algorithms: 1) the attack magnitude $\epsilon$ (also equivalent to the max $l_{p}$-norm of the perturbation) and 2) optimization steps $t$ for the attack. To limit the search scope of two hyper-parameters, two intervals are given: $\epsilon \in [0, \epsilon_{max}]$ and $t \in [0,t_{max}]$, where $\epsilon_{max}$ and $t_{max}$ are the max magnitude and iteration of each attack predefined by users. In this paper, we do not search for the attack step size, as it is relevant to optimization step $t$. Instead, all attacks 
that require step size parameter ({\it e.g.}, PGD) are modified to step size free version based on previous method~\cite{croce2020reliable}. Accordingly, the step size can be changed adaptively based on the optimization step. Then we can define the policy $s$ as the composition of various attacks, which consists of $N$ consecutive attack operations:
\begin{equation}
\label{Eq:2}
    s: \mathcal{A}^{s}_{N}(\mathcal{A}^{s}_{2}(\mathcal{A}^{s}_{1}(x,\mathcal{F};\epsilon_{s_{1}},t_{s_{1}}),\mathcal{F};\epsilon_{s_{2}},t_{s_{2}}))...,\mathcal{F};\epsilon_{s_{N}},t_{s_{N}}),
\end{equation}
where $\left \{ \mathcal{A}^{s}_{n} \in \mathbb{A}|  n=1,...,N \right \}$ is the sampled attacker from $\mathbb{A}$ separately and $\left \{ \{\epsilon_{s_{n}}, t_{s_{n}}\} |  n=1,...,N \right \}$ is the hyper-parameter of each attack. With the combination of different attack operations and hyper-parameters, we can obtain thousands of possible policies.

\begin{figure}[!htb]
\centering
\includegraphics[width=8.5cm]{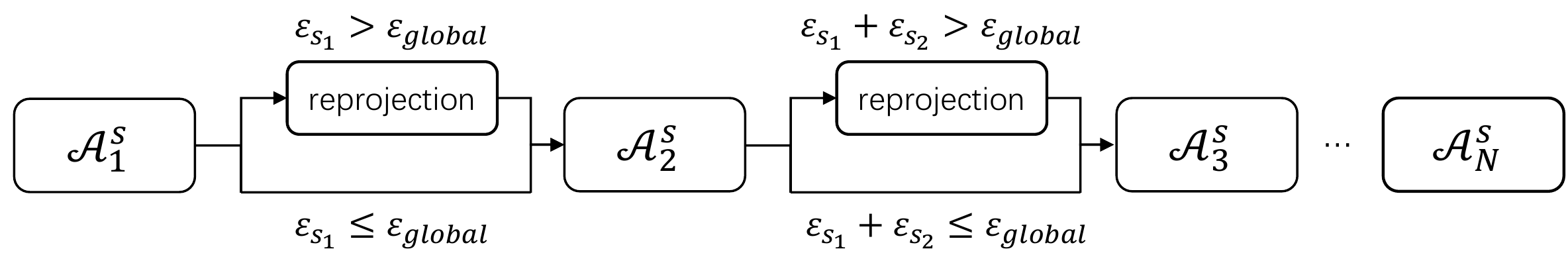}
\caption{Illustration of re-projection module in composite adversarial attack under $l_{p}$-norm constraint.}
\label{fig:intro3}
\end{figure}

\subsection{Constraining $l_{p}$-Norm by Re-projection}
The attack policy presented in Eq.~\ref{Eq:2} is a general form, which has no constraint on the global perturbation. When the attack sequence becomes longer, the computed perturbation of each attack algorithm is accumulated, causing the final perturbation on the original input to be large. To solve this problem, we insert a re-projection module between two consecutive attack algorithms. In Fig.~\ref{fig:intro3}, the re-projection module first determines whether the $\epsilon$ accumulated on previous attackers is larger than the $\epsilon_{global}$ of the policy. If it is, the accumulated perturbation will be clipped or rescaled to make the $l_{p}$-norm bounded in $\epsilon_{global}$. With this modification, we can use composite adversarial attacks for any $l_{p}$-norm conditions.

\subsection{Search Objective}
Previous works commonly use Attack Success Rate (ASR) or the Robust Accuracy (RA) as the objective to design their algorithms. However, these objectives can be achieved at the expense of more elapsed time. For example, recent proposed works~\cite{gowal2019alternative, tashiro2020output} use some tricks such as random restart or multiple targets to get higher success rate, with sacrificing the running efficiency. It makes their algorithms 
extremely slow (even more time-consuming than some black-box attacks). In this work, we emphasize that a good and strong attacker should be both effective and efficient. To meet this goal, we design our objective with minimizing two terms, {\it i.e.}, robust accuracy and complexity.

Next we elaborate the two objective terms. The first term RA is the accuracy of the target model on generated adversarial examples. It also reflects the strength of the attackers. As for the second term complexity, we use the number of gradient evaluation as the complexity metric. For a regular attack algorithm, the number of gradient evaluation represents the number of times an attack algorithm computing the gradient of the target model during the attack process, and it equals to the optimization step $t$ typically. Therefore, we can formulate the overall objective function as:
\begin{equation}
\label{Eq:3}
    \mathcal{L} = -\sum_{x}\left [\mathcal{F}(s(x))\neq y\right ]+ \alpha \sum_{i=0}^{N} t_{s_{i}},
\end{equation} 
where $s(x)$ represents the output of the attack policy for input $x$, $N$ is the length of the attack policy, and $\alpha$ is a coefficient to trade-off the attack strength and complexity. Then, we can apply a search algorithm to find an optimal attack policy $s^{\ast}$ from thousands of possible policies, by minimizing the objective $\mathcal{L}$:

\begin{figure}[!htb]
\centering
\includegraphics[width=8.50cm]{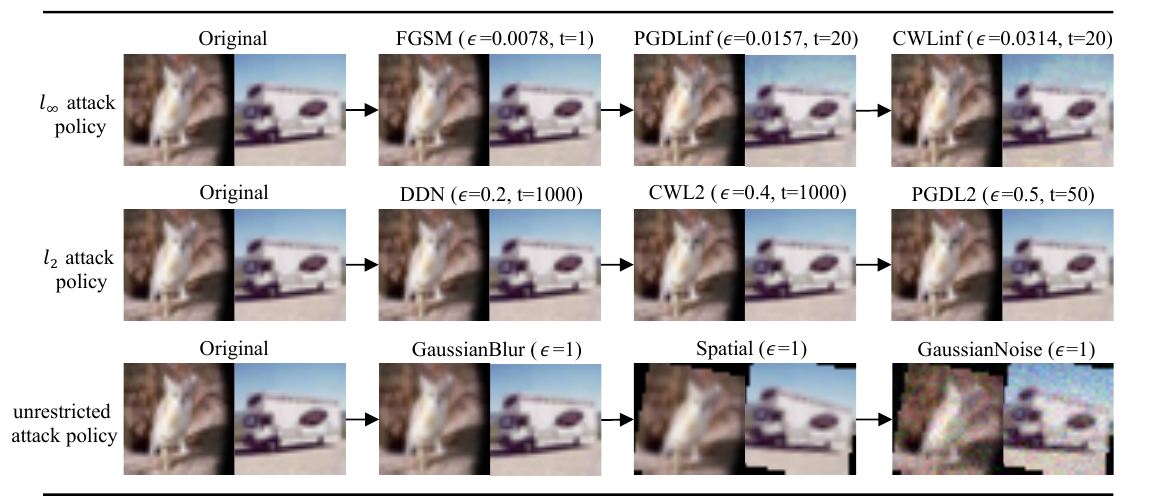}

\caption{Output visualization of an specific attack policy. For instance, the first row is an $l_{\infty}$ policy consists of consecutive FGSM, PGD-Linf and CW-Linf attack. Each column presents the output of each attacker component. For unrestricted attacks, $\epsilon$ is unlimited, so we set $\epsilon=1$.}
\label{fig:method3}
\end{figure}

\begin{equation}
\label{Eq:4}
    s^{\ast}=\min_{s} \mathcal{L}.
\end{equation}

\subsection{Search Space}
Our search space is divided into two parts: 1) searching for choices and orders of attack operations; 2) searching for magnitudes $\epsilon$ and steps $t$ of each attack operation. For an attack policy consisting of $N$ base attack operations, the attack operations search forms a problem space of $\left \|\mathbb{A} \right \|^{N}$ possibilities. Additionally, each operation is also associated with their magnitude and step. We discretize the range of magnitudes $\epsilon$ and steps $t$ into 8 values (uniform spacing) so that we can simplify the composite adversarial attacks search as a discrete optimization problem. Finally, the overall search space has total size of $(8\times8\times\left \|\mathbb{A} \right \|)^{N}$. 

In this paper, three types of policy spaces, {\it i.e.}, $S_{l_{\infty }}$, $S_{l_{2}}$ and
$S_{unrestricted}$ are constructed. We implement six $l_{\infty }$-attackers and six $l_{2}$-attackers in space $S_{l_{\infty }}$ and $S_{l_{2}}$ respectively. In unrestricted case, we use a larger search space with 19 implemented attack algorithms. Besides, all of $S_{l_{\infty }}$, $S_{l_{2}}$ and $S_{unrestricted}$ also adopt an IdentityAttack to represent the identity operation. An output visualization of attack policy in each search space is shown in Fig.~\ref{fig:method3}.

\subsection{Search Strategy}
The search strategy plays an important role in finding the best attack policy. In our problem setting, the scale of search space is relatively small. And the cost of policy evaluation is much less than other task such as NAS. This allows us to use some high-performance search algorithms. We compare three widely used methods, {\it i.e.}, Bayesian Optimization~\cite{snoek2012practical}, Reinforcement Learning~\cite{zoph2016neural} and NSGA-II Genetic Algorithm~\cite{deb2002fast}. The detailed implementation and comparison are shown in Appendix B. Although Bayesian Optimization and Reinforcement Learning are widely considered to be efficient in the field of AutoML, in this problem, we found that they are more time-consuming and slow to converge. In contrast, NSGA-II is faster since there is no need for an additional model optimization processes during the search period. It only needs a few iterations of the population updating to find an optimal solution quickly.

Detailly, NSGA-II needs to maintain a finite set $S$ of all possible policies and a policy evaluation function that maps each policy $s \in S$ onto the set of real numbers $\mathbb{R}$. In this work, we use Eq.~\ref{Eq:3} as the policy evaluation function. NSGA-II algorithm explores a space of potential attack policies in three steps, namely, a population initialization step that is generating a population $P_{0}$ with random policies, an exploration step comprising crossover and mutation of attack policy, and finally an exploitation step that utilizes the hidden useful knowledge stored in the entire history of evaluated policies and find the optimal one. The whole process is shown in Alg.~\ref{alg:adv_nsga}. In the remain of this work, we adopt NSGA-II algorithm for the policy search.

\begin{algorithm}[t!]
  \caption{\label{alg:adv_nsga} Attack policy search using NSGA-II}
  \begin{algorithmic}[1]
    \Require{Pool of candidate attackers $\mathbb{A}$; Population size $P$;}
    \Require{Maximum number of generations $G$}
    \State $P_{0} \gets \O $ \Comment{Initialized population with size of $K$} 
    \State $t \gets 0$
    \For{$i \gets 1$ to $K$}
    \For{$j \gets 1$ to $N$}
    \State Random sample $\mathcal{A}_{j}$ from $\mathbb{A}$
    \State Random sample $\epsilon_{j} \sim [0,\epsilon_{max}] $, $t_{j} \sim [0,t_{max}]$
    \EndFor
    \State $s \gets \mathcal{A}_{N}(\mathcal{A}_{1}(x,\mathcal{F};\epsilon_{1}, t_{1})...),\mathcal{F};\epsilon_{N}, t_{N})$
    \State $P_{0} \gets P_{0}\cup s$
    \EndFor
    \For{$t < G$} \Comment{Run search with Eq.~\ref{Eq:3} for evaluation}
    \State $P_{t+1} \gets $ NSGA-II($P_{t}$) \Comment{Update the populations}
    \State $t \gets t+1$
    \EndFor\\
\Return best attack policy $s^{\ast}$ from $P_{t}$
  \end{algorithmic}  
\end{algorithm}
%%%%%%%%%%%%%%%%%%%%%%%%%%%%%%%%%%%%%%%%%%%%%%%%%

\section{Experiments}
\subsection{Experiment Setup}
In order to validate the performance of our CAA, the searched attack policies on $S_{l_{\infty}}$, $S_{l_{2}}$ and $S_{unrestricted}$ are evaluated on 11 open source defense models. We run $l_{\infty}$ and $l_{2}$ attack experiments on CIFAR-10~\cite{krizhevsky2009learning} and ImageNet~\cite{deng2009imagenet} datasets. We perform unrestricted attack on Bird\&Bicycle~\cite{brown2018unrestricted} datasets. The robust accuracy is recorded as measurement to make comparison with 10 recent top attackers. In the implementation, we take all intermediate results of the policy and ensemble them as similar to \cite{croce2020reliable}.

\subsubsection{Details of Search Space.} The candidate pool of CAA consists of 32 attack operations, {\it i.e.}, six $l_{\infty}$-attacks, six $l_{2}$-attacks, 19 unrestricted attacks and the last IdentityAttack ({\it i.e.}, identity operation). A detailed summary of implemented attack algorithms is shown in Tab.~\ref{tab:visattack}. We borrowed the codes of some algorithms from open source attack toolbox, such as Foolbox~\cite{rauber2017foolbox} and Advertorch~\cite{ding2019advertorch}. The implementation and citation of each base attacker can be found in Appendix A.

\begin{table}[h]\small
\caption{The implemented attack algorithms in search space $S_{l_{\infty}}$, $S_{l_{2}}$, $S_{unrestricted}$ respectively.}
\label{tab:visattack}
\begin{center} 
\begin{tabular}{|c|c|c|}
\hline
$S_{l_{\infty}}$                                                                                                              & $S_{l_{2}}$                                                                                                        & $S_{unrestricted}$                                                                                                         \\ \hline
\begin{tabular}[c]{@{}c@{}}MI-LinfAttack\\ MT-LinfAttack\\ FGSMAttack\\ PGD-LinfAttack\\ CW-LinfAttack \\
SPSAAttack \\
IdentityAttack\end{tabular} & \begin{tabular}[c]{@{}c@{}}DDNAttack\\ CW-L2Attack\\ MI-L2Attack\\ PGD-L2Attack\\ MT-L2Attack \\
SquareAttack \\
IdentityAttack\end{tabular} & \begin{tabular}[c]{@{}c@{}}17 CorruptionAttacks\\ SpatialAttack\\ SPSAAttack \\
IdentityAttack\end{tabular} \\ \hline
\end{tabular}
\end{center}
\end{table}

\subsubsection{Data configuration}
For CIFAR-10, we search for the best policies on a small subset, which contains 4,000 examples randomly chosen from the train set. Total 10,000 examples in test set are used for the evaluation of the searched policy. For ImageNet, as the whole validation set is large, we randomly select 1000 images for policy search and 1000 images for evaluation from training and testing database respectively. For Bird\&Bicycle, we use all 250 test images for evaluation, and 1000 randomly selected training images for attack policy search.

\subsubsection{Summary of Experiments}
We investigate four cases: 1) BestAttack, searching for best single attacker in candidate pool; 2) EnsAttack, searching for the ensemble of multiple attackers; 3) CAA$_{dic}$, directly searching CAA policy on given datasets; and 4) CAA$_{sub}$, searching by attacking adversarial training CIFAR10 model as substitute and transferred to other models or tasks. For fairness, we compare our method with previous state-of-the-art attackers on 11 collected defense models: Advtrain~\cite{madry2017towards}, TRADES~\cite{zhang2019theoretically}, AdvPT~\cite{hendrycks2019using}, MMA~\cite{ding2019mma}, JEM~\cite{grathwohl2019your}, PCL~\cite{mustafa2019adversarial}, Semi-Adv~\cite{carmon2019unlabeled}, FD~\cite{xie2019feature}, AdvFree~\cite{shafahi2019adversarial}, TRADESv2\footnote{https://github.com/google/unrestricted-adversarial-examples} and LLR\footnote{https://github.com/deepmind/deepmind-research/tree/master/ unrestricted\_advx}. Next we leverage multiple architectures (VGG16~\cite{simonyan2014very}, ResNet50~\cite{he2016deep}, Inception~\cite{szegedy2015going}) and datasets (MNIST~\cite{lecun1998gradient}, CIFAR-100, SVHN~\cite{netzerreading}) to investigate the transferability of CAA in black-box and white-box settings. Finally, we do some ablation study on the effect of different policy search algorithms and the attack policy length $N$. We also analyse the difference between searched policies of non-target and target attack. Some insights can be found in these ablation experiments.

% Next, we will study the transferability of AutoAttack from two aspects: 1) the searched attack policy cross tasks or models and 2) the attack itself under black box transfer attack settings. For 1), we search the policy by attacking CIFAR-10 classification trained on four different architectures: VGG16, Inception, ResNet and compare the differences. For 2),
% we use a substitutive model in the attack phase. In this case, target model in attack phase and policy search phase are totally different. Such a modification encourage the optimization algorithm to search a more transferable attack policy. The results also show that attack policy are surprisingly transferable and yield improvements even on different models or tasks. And AutoAttack also gains the promotion on black box substitution model transfer attack setting.

% Finally, we do some ablations about the effect of different policy search algorithms, the attack policy length $N$. We also analyse the difference of searched policies in non-target and target attack, find some insights to design attack algorithms in non-target or target setting.

\subsection{Comparison with State-of-the-Art}
\label{sec:exp1}

\begin{table*}[t]
		\label{tab:policy_vis}
	\centering
		%\begin{small}
		%\centering
		\small
		\begin{tabular}{l || c}
	    	\specialrule{0.1em}{3pt}{3pt}
	    	\multicolumn{2}{c}{\textbf{Visualization of CAA$_{sub}$ proxy attack policies }} \\
	    	\specialrule{0.1em}{3pt}{3pt}
			$S_{l_{\infty}}$ & [('MT-LinfAttack', $\epsilon$=8/255, $t$=50), ('MT-LinfAttack', $\epsilon$=8/255, $t$=25), ('CWLinfAttack', $\epsilon$=8/255, $t$=125)] \\
			$S_{l_{2}}$ & [('MT-L2Attack', $\epsilon$=0.5, $t$=100), ('PGD-L2Attack', $\epsilon$=0.4375, $t$=125), ('DDNAttack', $t$=1000)] \\
			$S_{unrestricted}$ & [('FogAttack', $\epsilon=1$, $t=1$), ('FogAttack', $\epsilon=1$, $t=1$), ('SPSAAttack', $\epsilon$=16/255, $t$=100)] \\
		\specialrule{0.1em}{3pt}{3pt}
	\end{tabular}\end{table*}

\begin{table*}[t] 
\vspace{-3mm}
	\centering
		%\begin{small}
		%\centering
		\small
		\begin{tabular}{l || cccccccc}
	    	\specialrule{0.1em}{3pt}{3pt}
			\textbf{CIFAR-10} - $l_\infty$ - $\epsilon=8/255$ & AdvTrain & TRADES & AdvPT & MMA & JEM & PCL & Semi-Adv & Complexity \\ 
	\specialrule{0.1em}{3pt}{3pt}
			 PGD~\cite{madry2017towards} & 51.95 & 53.47 & 57.21 &50.04 & 9.21 & 8.12 & 61.83 & 1000 \\
			 FAB~\cite{croce2019minimally} & 49.81 & 51.70 & 55.27 &42.47 &62.71 &0.71 & 60.12 & 1350 \\
			 APGD~\cite{croce2020reliable} & 51.27 & 53.25 &  56.76 &49.88 & 9.06 &7.96 & 61.29 & 1000 \\
			 AA~\cite{croce2020reliable} & 49.25 & 51.28 & 54.92 & 41.44 & 8.15 &0.28 & 59.53 & 4850 \\
			 ODI-PGD~\cite{tashiro2020output} & 49.37 & 51.29 & 54.94 & 41.75 &8.62 &0.53 & 59.61 & 3200 \\			BestAttack on $S_{l_{\infty}}$ & 50.12 &52.01 & 55.23 & 41.85 &9.12 &0.84 & 60.74 & 900 \\
			 EnsAttack on $S_{l_{\infty}}$ & 49.58 & 51.51 & 55.02 &41.56 &8.33 &0.73 & 60.12 & 800 \\
			 \rowcolor[gray]{0.9}
			 CAA$_{sub}$ on $S_{l_{\infty}}$ & 49.18 & 51.19 & 54.82 &40.87 & 7.47 & 0.0 & 59.45 & \textbf{800} \\
			 \rowcolor[gray]{0.9}
			 CAA$_{dic}$ on $S_{l_{\infty}}$ & \textbf{49.18} & \textbf{51.10} & \textbf{54.69} & \textbf{40.69} &\textbf{7.28} &\textbf{0.0} & \textbf{59.38} & - \\
			 \specialrule{0.1em}{3pt}{3pt}
	\end{tabular}
	\caption{The table is divided into two parts. The lower part presents the reported RA(\%) of $l_{\infty}$-based attack on diverse CIFAR-10 defenses. Each column presents the result on a specific defense and the last column presents the complexity of the attack algorithm. The upper part of the table presents the best attack policies found by our method.}
	\label{tab:exp_cifar1}
	\end{table*}

Tab.~\ref{tab:exp_cifar1} shows the $l_{\infty}$-based attack result of four variants, {\it i.e.}, CAA$_{sub}$, CAA$_{dic}$, EnsAttack and BestAttack on CIFAR-10 dataset. Most works study the model robustness in this setting, so we can collect more defenses for evaluation. The compared attackers are 150-step ODI-PGD with 10 ODI-step and 20 restarts, 100-step PGD\&APGD with 10 restarts, FAB and AA. The hyper-parameters of FAB and AA are consistent with the original paper~\cite{croce2020reliable}. All these attackers have the total number of gradient evaluation (complexity) larger than 1000. In contrast, our CAA$_{sub}$ has lower complexity (800), and breaks the model with a higher error rate. It implies that even a substitute attack policy may have high time efficiency and reliability. Direct search on the task of interest can further improve the performance. From the last row of the table, we can see stronger attack policies founded by CAA$_{dic}$, with the average decrease of 0.1\% on the robust accuracy. Except for CAA, we also evaluate two optional schemes in Fig.~\ref{fig:intro2}, named BestAttack and EnsAttack. The final searched policy of BestAttack is MT-LinfAttack, which is the strongest attacker in $S_{l_{\infty}}$ case. However, the result shows the best single attacker is not competitive in front of existing methods. EnsAttack searches a policy with an ensemble of MT-Linf, PGD-Linf and CW-Linf attacks. Compared to BestAttack, EnsAttack 
merges multiple attacks and achieves better results. But it is still worse than CAA policy. It implies that CAA are empirically better than ensemble of attackers. For $l_{2}$-based attack on CIFAR-10, our method also yields excellent performance. 

% The the best performed baseline at present is AA [], which is the ensemble of four strong attacks, namely APGD-CE, APGD-DLR, FAB and Square attacks. Although AA achieves impressive 49.25\% accuracy on adversarial training models (4\% lower than PGD), it takes a lot of computation and time to optimize a batch of perturbation (3600 forward-backward on target model). In contrast, our AutoAttack$_{sub}$ achieves slightly better performance than AA, and meanwhile greatly reduces the elapsed time 
% and resources cost, with only 500 optimization steps. From the table we can see Ens-MultiAttacker is wearker than AutoAttack. It suggested that forming a sequence of attackers instead of ensemble is reasonable, which yields an improvement of around 1\%. Another improvement can be obtained by direct searching the attack policy on target models and datasets. It may cost more time, but we show that the search time in this task is relatively small than some automation technologies such as NAS or AutoAugment. A detailed search time analysis is in sec [].

The result on ImageNet is shown in Tab.~\ref{tab:exp_2}. We show that CAA gains greater improvement on ImageNet, compared to CIFAR-10. In particular, CAA$_{sub}$ achieves 38.30\% accuracy attacking $l_{\infty}$ adversarially trained models, with around 2\% improvement over state-of-the-art. It implies that CAA is more suitable for attacking complex classification tasks. ImageNet classification has more categories and larger image input size. Also, we found the adversarial examples generated by base attackers are more diverse on ImageNet. For such a complex task, there is more room for the attack strategy design.

For unrestricted attack, we choose the benchmark of Bird\&Bicycle proposed in \textit{unrestricted adversarial examples contest}~\cite{brown2018unrestricted}. The top two defense models LLR and TRADESv2 on leaderboard are used for evaluation. 
For fairness, we only use warm-up attacks in contest as our search space $S_{unrestricted}$, and avoid the attacks that the defense model has never seen before. Both LLR and TRADESv2 get nearly 100\% robust accuracy on Corruption, Spatial and SPSA attacks. But after composing these attacks by CAA, the robust accuracy of LLR and TRADESv2 is rapidly dropped to around zero. The result shows that existing unrestricted adversarial defense models are severely overfitting to the single test attackers. In unrestricted attack setting, there is no good defense against our CAA. Therefore, we think there is still a lot of work to do for achieving the truly unrestricted adversarial robustness.

\subsubsection{Analysis of searched policy}
We visualize the searched best policy on $S_{l_{\infty}}$, $S_{l_{2}}$ and $S_{unrestricted}$ in Tab.~\ref{tab:exp_cifar1}. The presented policy is searched by attacking adversarially trained model on CIFAR-10 classification task. In all $l_{\infty}$, $l_{2}$ and unrestricted attack scenarios, CAA tends to choose strong attacks. Take policy of $S_{l_{\infty}}$ as an example, CAA chooses the strongest MT-LinfAttack as the first and the second position attack, and abandons the weaker attackers, such as one-step FGSM. Therefore, we think a well selected candidate attacker pool is critical to the performance of CAA. Another foundation is that CAA prefers some policies with the combination of diverse base attackers. It means that a policy formed with MI-Linf and PGD-Linf attack always yields little improve, because the difference among them are subtle (with the
same principle and objective function). In contrast, in the best policy of $S_{l_{\infty}}$, CAA selected a more diverse margin loss based CW-Linf attack to assist cross entropy loss based attackers, which promotes the attack performance. 

\begin{table}[t] \caption{RA (\%) under $l_{2}$ and unrestricted attacks, experimented on ImageNet, CIFAR-10 and Bird\&Bicycle datasets.}
		\label{tab:exp_2}
	\centering
		%\begin{small}
		%\centering
		\small
		\begin{tabular}{l ||cc}
	    	\specialrule{0.1em}{3pt}{3pt}
			\textbf{CIFAR-10} - $l_{2}$ - $\epsilon=0.5$ & AdvTrain & MMA \\
		\specialrule{0.1em}{3pt}{3pt}
		DDN~\cite{rony2019decoupling}&  69.86 & 66.21 \\
		    FAB~\cite{croce2019minimally}&
			69.46 & 66.33 \\
			AA~\cite{croce2020reliable}&
			69.26 & 66.09 \\
			CAA$_{sub}$ on $S_{l_{2}}$ &
			69.22 & 65.98 \\
			CAA$_{dic}$ on $S_{l_{2}}$ &
			\textbf{69.20} & \textbf{65.95} \\
			
			 \specialrule{0.1em}{3pt}{3pt}
			\textbf{ImageNet} - $l_{2}$ - $\epsilon=3$ & AdvTrain & AdvFree \\
		\specialrule{0.1em}{3pt}{3pt}
			  DDN~\cite{rony2019decoupling}&  38.1 & 34.65 \\
		    FAB~\cite{croce2019minimally}&
			36.93 & 34.46 \\
			AA~\cite{croce2020reliable}&
			36.3 & 34.11 \\
			CAA$_{sub}$ on $S_{l_{2}}$ & 
			35.18 & 33.95 \\
			CAA$_{dic}$ on $S_{l_{2}}$ &
			\textbf{35.07} & \textbf{33.89} \\
			  
			  \specialrule{0.1em}{3pt}{3pt}
			\textbf{ImageNet} - $l_{\infty}$ - $\epsilon=4/255$ & AdvTrain & FD \\
		\specialrule{0.1em}{3pt}{3pt}
		    APGD~\cite{croce2020reliable}&  42.87 & 23.18 \\
		    FAB~\cite{croce2019minimally}&
			41.24 & 21.86 \\
			AA~\cite{croce2020reliable}&
			40.03 & 21.54 \\
			CAA$_{sub}$ on $S_{l_{\infty}}$ &
			38.30 & 19.41 \\
			CAA$_{dic}$ on $S_{l_{\infty}}$ &
			\textbf{38.21} & \textbf{19.27} \\
			 \specialrule{0.1em}{3pt}{3pt}
			\textbf{Bird\&Bicycle - $unrestricted$} & LLR & TRADESv2 \\
			\specialrule{0.1em}{3pt}{3pt}
			Common Corruptions &  100.0 & 100.0  \\
		    Spatial~\cite{engstrom2019exploring}&  100.0 & 99.5 \\
		    Boundary~\cite{brendel2017decision}&
			100.0 & 95.0 \\
			SPSA~\cite{uesato2018adversarial}&
			100.0 & 100.0 \\
			CAA$_{dic}$ on $S_{unrestricted}$ &
			\textbf{7.9} & \textbf{4.0} \\

		\specialrule{0.1em}{3pt}{3pt}

	\end{tabular}\end{table}

\subsection{Attack transferability}
We study the transferability of CAA in two scenarios: 1) black-box setting and 2) white-box setting. In black-box setting, we cannot obtain the gradient of the target model. Instead, we use CAA to search a policy on substitute model and generate adversarial examples to attack the target model. In white-box setting, gradient evaluation is allowed, so policy searched on substitute tasks or models are used for generating adversarial examples directly on the target model. 

% It contains two aspects: 1) the transferability of searched attack policy cross tasks or models and 2) the transferability under black box transfer attack using substitute model.

\subsubsection{Black-box Transferability of CAA}
Here we discuss if CAA can be used for searching black-box transfer attacks. We slightly modify the original CAA to meet this requirement. Specifically, we use an attack policy $s$ to attack substitute model at adversarial example generation stage. Then these adversarial examples is tested on target model. The robust accuracy on target model is regarded as the evaluation score of the policy $s$. Except for this, the entire search process remains unchanged. We name this variation as CAA$_{trans}$. In the attack transferability experiment, we use three types of models (VGG16, Inceptionv3 and ResNet50) with different architectures, and all of them are defended by standard adversarial training. The result is recorded in Tab.~\ref{tab:exp_3}. The first column presents the experiment setting. For example, R$\rightarrow$V means that we use ResNet50 as substitute model to attack VGG16. 

We show that CAA gains better performance in most transfer attack settings. Especially, it significantly increases the attack strength when VGG16 is used for substitute model, causing the decrease of 3\% on target model accuracy. The result suggests that an automatic search process also helps for discovering a more black-box transferable attack policy, not limited to white box scenarios. From the visualization of searched transferable policy in Appendix D, we found that CAA$_{trans}$ does not adopt some ``strong" attacks, since such attacks may have poor transferability. Oppositely, attacks like FGSM or MI-Linf attack are chosen as better transferable component in the policy, which explains why CAA$_{trans}$ could improve the attack transferability.

\begin{table}[t] \caption{Black-box transfer attack results on CIFAR-10. R, V and I represent ResNet, VGG and Inception respectively.}
		\label{tab:exp_3}
	\centering
		%\begin{small}
		%\centering
		\small
		\begin{tabular}{l ||ccc}
			 \specialrule{0.1em}{3pt}{3pt}
			\textbf{Models} & BestAttack & EnsAttack & CAA$_{dic}$ \\
			\specialrule{0.1em}{3pt}{3pt}
		    R $\rightarrow$ V& 64.75 & 63.93 & \textbf{63.85} \\
		    R $\rightarrow$ I & 67.34 &
			\textbf{66.05} & 66.21 \\
			V $\rightarrow$ R & 67.21 & 65.23 &  \textbf{64.98} \\
			V $\rightarrow$ I & 63.42 & 61.33 & \textbf{60.81} \\
			I $\rightarrow$ R & 65.29 & 64.98 & \textbf{64.38} \\
			I $\rightarrow$ V & 59.82 & 58.54 & \textbf{58.32} \\
		\specialrule{0.1em}{3pt}{3pt}
	\end{tabular}\end{table}

% \begin{table}[t] \caption{Todo: transfer}
% 		\label{tab:eval_additional_defenses}
% 	\vspace{2mm}
% 	\centering
% 		%\begin{small}
% 		%\centering
% 		\small
% 		\begin{tabular}{l || R{10mm}{R{10mm}R{10mm}} >{\columncolor[rgb]{0.9 1.0 0.9}}}
% 			 \specialrule{0.1em}{3pt}{3pt}
% 			\textbf{Models} & Baseline & AutoAttack$_{sub}$ & AutoAttack$_{dic}$ \\
% 			\specialrule{0.1em}{3pt}{3pt}
% 		    VGG16 &
% 			42.41 & 41.98 & 41.73 \\
% 			DenseNet &
% 			50.07 & 48.93 & 48.77\\
% 			Inception &
% 			48.12 & 46.80 & 46.61\\
% 		\specialrule{0.1em}{3pt}{3pt}
% 		\textbf{Datasets} & Baseline & AutoAttack$_{sub}$ & AutoAttack$_{dic}$ \\
% 		\specialrule{0.1em}{3pt}{3pt}
% 					SVHN &
% 			56.36 & 54.91 & 54.74\\
% 					MNIST &
% 			88.61 & 88.32 & 88.18\\
% 					CIFAR-100 &
% 			17.84 & 16.77 & 16.54 \\
% 			\specialrule{0.1em}{3pt}{3pt}
% 	\end{tabular}\end{table}
	
\begin{table}[t] \caption{Comparison of different optimization algorithms for the attack policy search.}
		\label{tab:exp_4}
	\centering
		%\begin{small}
		%\centering
		\small
		\begin{tabular}{l ||ccc}
			\specialrule{0.1em}{3pt}{3pt}
			\textbf{Search Methods} & Performance & Search time \\
			\specialrule{0.1em}{3pt}{3pt}
		    Random Search-100 & 52.09 & 8 Hours \\
		    Reinforcement Learning &
			51.44 & 5 GPU/d \\
		    Bayesian Optimization &
			50.02 & 5 GPU/d \\
			NSGA-II &
			49.18 & 3 GPU/d & \\
			\specialrule{0.1em}{3pt}{3pt}
	\end{tabular}\end{table}
	
\subsubsection{White-box Transferability of CAA}
Here we seek to understand if it is possible to transfer attack policies in white-box case, namely, policies searched on substitute tasks or models are used for attacking the target model. A detailed experiment is presented in Appendix C.
From the result, we highlight that the searched policies on CIFAR-10 still transfer well to many model architectures and datasets. Therefore, we believe that CAA does not ``overfit” to the datasets or model architectures and it indeed finds effective policies that catch the true weakness and can be applied to all kinds of such problems. However, there is no guarantee that the attack policies are transferred across defenses. One empirical practice to improve the transferability across defenses is using stronger and more diverse attack algorithms in candidate pool. A reference is in Tab.~\ref{tab:exp_cifar1}, by using six strong attackers in $S_{l_{\infty}}$, CAA$_{sub}$ has achieved satisfactory results on multiple defense models.

% To validate, we experiment AutoAttack on diverse models and datasets (as shown in Table []). Compared to baseline, CAA policies are never found to hurt the performance of models
% even if they are learned on a different datasets or architectures. For example, the policy learned to attack ResNet50 for CIFAR-10 task leads to the improvements to attack all of the other model architectures trained on full CIFAR-10. Similarly, a policy learned on attacking CIFAR-10 task also works for other datasets that have different data and class distributions. Therefore, we believe that AutoAttack does not “overfit” to the datasets or models and it indeed finds efficient policies that catch the true weakness and can be applied to all kinds of such problems. Although learning policies directly on given models and datasets gets better results, the performance gap is so small or even negligible. At this case, we do not think searching from scratch is necessary.

\subsection{Ablations}
\subsubsection{Analysis of the policy length $N$}
\begin{figure}[htbp]
\centering

\includegraphics[width=8.3cm]{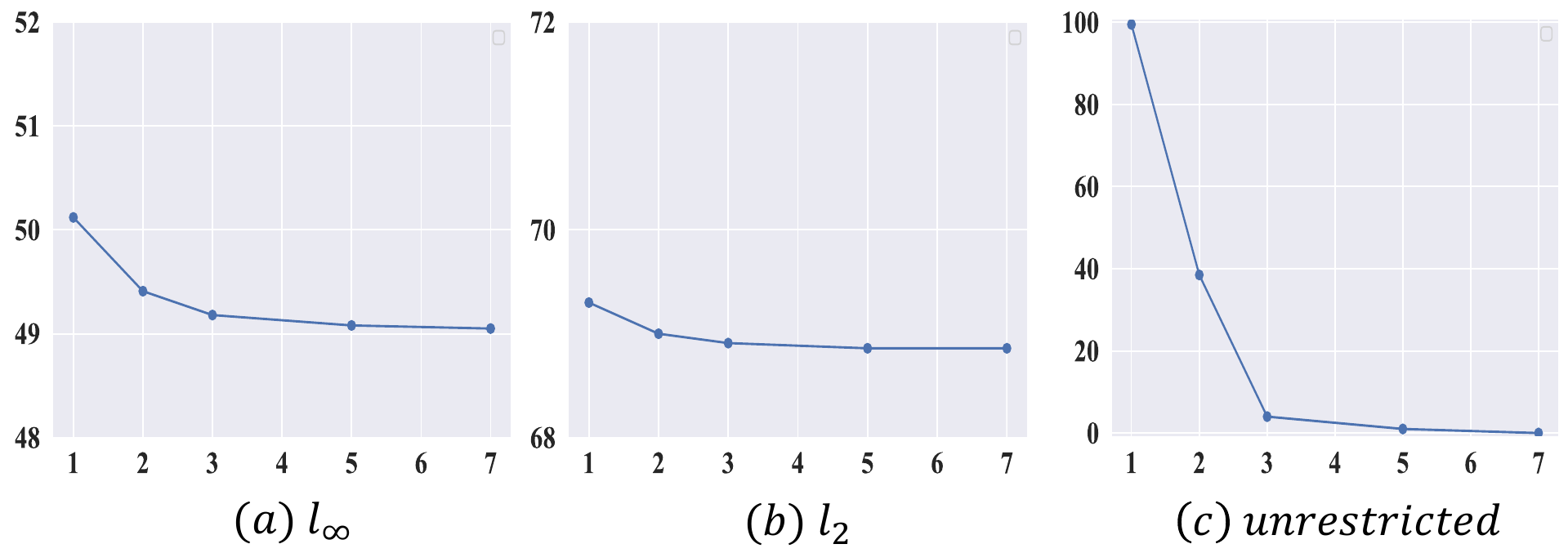}

\caption{Effect of the policy length on attack performance in $l_{\infty}$, $l_{2}$ and unrestricted settings.}
\label{fig:exp_1}
\end{figure}

We conduct a series of experiments to explore if a longer policy, which can adopt more and diverse base attackers, exhibits stronger attack ability. We choose five policies with length of 1, 2, 3, 5 and 7. Fig.~\ref{fig:exp_1} shows the curve of robust accuracy with the policy length. The CAA equals to find the best base attacker in the candidate pool when $N=1$, so that the performance is the worst in this case. With the increase of $N$, the attack policy becomes stronger in all $l_{\infty}$, $l_{2}$ and unrestricted settings. We found that the policy length has the smallest effect on $l_{2}$ attack settings. It is reasonable that more base attack just means more optimization steps for $l_{2}$ attack. In contrast, $N$ greatly influences the performance on unrestricted attack. The accuracy quickly drops to around zero in unrestricted setting when using a searched attack policy larger than 3.  

\subsubsection{Different Search Methods}
Tab.~\ref{tab:exp_4} presents the performance and search time of four optimization methods, {\it i.e.}, Random Search, Bayesian Optimization, Reinforcement Learning and NSGA-II Genetic Algorithm. The detailed implementation of each method is listed in Appendix B. Random Search, with 100 random 
policies for trials and the best one chosen, is regarded as a baseline. Compared to the baseline, all heuristic algorithms find better policies. Although Bayesian Optimization and Reinforcement Learning are widely considered to be efficient in searching of large space, in this problem, we found they are more time-consuming and prone to fall into local optimal. In contrast, NSGA-II finds better policies with the lower cost of 3 GPU/d, and achieves better performance.

\subsubsection{Target vs. Non-target attack}
Target attack is a special application scenario where attackers fool the model to output target label they wanted. Otherwise, it is called non-target attack that no target labels are given. We experiment our CAA under target attack settings in Appendix C. For target attack, CAA searches a policy with less random initialization. It indicates that attackers without a random initialization are more suitable for targeted setting. Also, compared to margin loss, base attackers with cross entropy loss are favoured by CAA. Policies searched by CAA also gains improvement in target attack.

\section{Conclusion}

We propose an automatic process of learning attack policies formed by a sequence of base attackers for breaking an ML system. By comparing our searched policy with 10 recent attackers on 11 diverse defense, we show that our method achieved better attack success rate with less running time. It empirically demonstrates that searching better algorithms and hyper-parameters also helps for the adversarial attacks.

We think the foremost extension of our work is how to defense attackers which can automatically search for the strongest attack algorithm. From this point of view, we are going to study the adversarial training method based on our CAA in future work.

\bibliography{bib.bib}

\clearpage
\appendix
  \renewcommand{\appendixname}{Appendix~\Alph{section}}

  \section{Appendix A: Implementation Details}
  \label{apdix:a}
  \subsection{Implementation of 32 base attackers}
  \subsubsection{FGSM Attack} FGSM~\cite{goodfellow2014explaining} performs a single step update on the original sample $x$ along the direction of the gradient of the loss function $\mathcal{L}(x, y; \mathcal{F})$. The loss function is usually defined as the cross-entropy between the output of the
network $\mathcal{F}$ and the true label $y$. Formally, FGSM adversarial
samples are generated as:
\begin{equation}
    x_{adv} = \text{clip}_{[0,1]}\{ x+\epsilon \cdot \text{sign}(\nabla_{x}\mathcal{L}(x, y; \mathcal{F})) \}
\end{equation}
where $\epsilon$ controls the maximum $l_{\infty}$ perturbation of the adversarial samples, and the $\text{clip}_{[0,1]}(x)$ function forces $x$ to reside in the range of $[0, 1]$. For FGSM, the $\epsilon$ is searchable and the step $t$ is fixed with 1.

\subsubsection{MI-Linf \& MI-L2 Attack} An iterative variant of FGSM is called Iterative FGSM attack. I-FGSM Attack can be expressed as:
    \begin{equation}
      x_{l+1}=\text{project}\{x_{l}+\epsilon_{step} \cdot \frac{\nabla_{x}\mathcal{L}(x_{l}, y; \mathcal{F})}{\left \| \nabla_{x}\mathcal{L}(x_{l}, y; \mathcal{F}) \right \|_{1}}\}
  \end{equation}
  where $\text{project}$ keeps $x$ within both predefined perturbation and the image value range. Furthermore, Momentum Iterative (MI) attack~\cite{dong2018boosting} is derived from the Iterative FGSM, which integrates the momentum term into an iterative process to generate adversarial samples. Given $g_{0}=0$ and $g_{l+1}=\mu \cdot g_{l}+(1-\mu) \cdot \frac{\nabla_{x}\mathcal{L}(x_{l}, y; \mathcal{F})}{\left \| \nabla_{x}\mathcal{L}(x_{l}, y; \mathcal{F}) \right \|_{1}}$, the MI attack can be expressed as:
  \begin{equation}
      x_{l+1}=x_{l}+\epsilon_{step} \cdot \text{sign}(g_{l+1})
  \end{equation}
  MI attack supports $l_{\infty}$ and $l_{2}$ settings according to different max $l_{p}$ norm. We set $\mu=0.25$ in the implementation.
  
  \subsubsection{PGD-Linf \& PGD-L2 Attack} Projected Gradient Descent (PGD)~\cite{madry2017towards} is
an universal adversary among all the first-order adversaries, which follows the update rule: 
  \begin{equation}
      x_{l+1}=\text{project}\{x_{l}+\epsilon_{step} \cdot \text{sign}(\nabla_{x}\mathcal{L}(x_{l}, y; \mathcal{F}))\}
  \end{equation}
  where $\text{project}$ keeps $x$ within both predefined perturbation range and the image value range. We enable the random starts of inputs since it can improve the diversity.

  \subsubsection{CW-Linf \& CW-L2 Attack ($l_{\infty}, l_{2}$)} Carlini \& Wagner (CW) attack~\cite{carlini2017towards} use a new loss function minimizing the distance in logit values between class $y$ and the second most-likely class:
  \begin{equation}
    \mathcal{L}_{CW}(x, y; \mathcal{F})= \max(\max_{i\neq y}(\mathcal{F}(x)_{(i)})-\mathcal{F}(x)_{(y)}, -\kappa )
\end{equation}
where $\kappa$ establish the margin of the logit of the adversarial class being larger than the logit of runner-up class. $\kappa$ presents a minimum desirable degree of robustness for the target adversary. For $l_{\infty}$ C\&W attacker, we directly replace the loss function $\mathcal{L}$ in PGD-Linf attack as $\mathcal{L}_{CW}$. For $l_{2}$ C\&W attacker, we optimize the $l_{2}$ perturbation as follows:
\begin{equation}
    x_{adv}=x+\text{argmax}_{\delta }(c\cdot\mathcal{L}_{CW}(x+\delta,y;\mathcal{F})+\left \|\delta \right \|_{2}^{2})
\end{equation}
where $c$ is the tradeoff coefficient with perturbation minimization term and classification error maximization term. We set $c=0.5$ and  $\kappa=20$ in our implementation. Adam with learning rate of 0.01 are used to optimize the perturbation.

\subsubsection{MT-Linf \& MT-L2 Attack} MultiTargeted Attack~\cite{gowal2019alternative} is an alternative of PGD attack. It uses several classes different with the true label $y$ as the target, and computes $r$ margin losses:
\begin{equation}
    \mathcal{L}^{(r)} = \mathcal{F}(x)_{(t^{(r)})}-\mathcal{F}(x)_{(y)} \  \text{with}\  t^{(r)}\in \mathcal{T}
\end{equation}
where $\mathcal{T}=\{1,2,...,C\}\backslash \{ y\}$ is the set which contains all classes other than $y$. $r$ is the number of the targets. In each restart of PGD attack, MultiTargeted Attack uses a surrogate loss with a different target, instead of the original cross entropy loss. In the paper, the authors give proofs that a regular PGD-based attack has no guarantee that it will find the global maximum. However, MultiTargeted Attack, which inspects all logit differences, is guaranteed to
find the global maximum. In the implementation, we set number of the targets as 9.

\subsubsection{DDN Attack} Decoupled Direction and Norm (DDN) Attack~\cite{rony2019decoupling} is a variant of CW $l_{2}$ attack, which decouples the perturbation
norm penalty and the direction which causes misclassification. Detailly, in each optimization step, DDN minimized the perturbation when $x_{adv}$ broke the model successfully, maximized the classification error when $x_{adv}$ cannot attack the model successfully. DDN converged faster and produced adversarial examples with higher perceptual similarity compared to the C\&W. In the implementation, we use SGD with learning rate 1.0 to optimize the perturbation. We adopt cosine annealing to schedule the learning rate. Part of 
the implementation code are refered to Advertorch.

\subsubsection{Square Attack} Square Attack~\cite{andriushchenko2019square} is a type of score-based black-box attack. It is based on a randomized search scheme which selects localized square shaped updates at random positions. The attack process is simple but efficient. As the most effective black box attack, Square Attack even outperforms white-box attacks in some attack scenarios. In the implementation, we set the initial size of squares $p$ as 0.8.

\subsubsection{Spatial Attack} Spatial Attack~\cite{engstrom2019exploring} is a type of unrestricted attack, which rotations and translations the input and significantly degrade the accuracy of image classifiers. Consistent with \cite{brown2018unrestricted}, we define the max translation offset on the X and Y axes as 5, the max degree of rotation as 30. Then a grid search is applied to find the spatial transform with the maximum loss value. We choose five possible values uniformly on translation and rotation, which leads to 125 possibilities. Finally, the 
largest-case on loss values is adopted as the result transformation.

\subsubsection{SPSA Attack}
Simultaneous Perturbation Stochastic Approximation (SPSA) Attack~\cite{uesato2018adversarial} is a numerical gradient estimation based black-box attack, which estimates the gradient using 2 queries to $\mathcal{F}(\cdot)$ by simultaneously perturbing all dimensions using a vector $v\sim \{-1,1\}^{D}$ sampled from the Rademacher distribution (\textit{i.e.}, whose elements are either $+1$ or $-1$):  
\begin{equation}
    \frac{\partial\mathcal{F}}{\partial x}\approx \frac{\mathcal{F}(x+\Delta v)-\mathcal{F}(x-\Delta v)}{2\Delta }\cdot v
\end{equation}
where $\Delta$ is a small
constant (\textit{e.g.}, $\Delta=10^{-9}$), $D$ is the dimensions of input image. In this work, SPSA attack is used for both $l_{\infty}$ and unrestricted setting. We set the learning rate as 0.01. For unrestricted attack setting, we fix the optimization step to 100. 

\subsubsection{17 Corruption Attacks} Except for worst-case perturbation, many forms of corruption such as snow, blur, pixelation, and novel combinations of these can also cause the misclassification of the model. Therefore, we adopt 17 kinds of image corruption operations in IMAGENET-C~\cite{hendrycks2019benchmarking}, each of them has the severity of 1. The detailed name of 17 corruptions are: $gaussian\_noise$, $shot\_noise$, $impulse\_noise$, $defocus\_blur$, $glass\_blur$, $motion\_blur$, $zoom\_blur$, $fog$, $brightness$, $contrast$, $elastic\_transform$, $pixelate$, $jpeg\_compression$, $speckle\_noise$, $gaussian\_blur$, $spatter$, $saturate$. Note that the unrestricted defenses (LLR, TRADESv2) used in this work are achieved nearly 100\% accuracy on all of corruptions. 

\subsubsection{Identity Attack} The last one is identity attack, which is the operation that do nothing to the input (\textit{i.e.}, $x_{adv}=x$). 
  
\subsection{Implementation of searching on $l_{\infty}$ Space}
  For $l_{\infty}$ attack policy search, we set the $\epsilon_{max}$ in each base attacker as the max epsilon $\epsilon_{global}$ of the whole attack policy. The $t_{max}$ is set separately based on different attack algorithms. For iterative gradient based method, we set $t_{max}=200$. For black-box methods, we set $t_{max}=1000$. To caculate the final onjective $\mathcal{L}$, we set $\alpha = 10e-4$. On the NSGA-II search algorithm, we set the initialized population size $K=20$, the maximum number of generations $G=40$ and the offsprings in each generation as the number of 10. The best policy are choosed by the lowest value computed by $\mathcal{L}$.

  \subsection{Implementation of searching on $l_{2}$ Space}
  The $l_{2}$ attack policy search is the mostly same with $l_{\infty}$. For $l_{2}$ attack, we set $t_{max}=2000$. Since there are more optimization steps in $l_{2}$ attacker, we set the tradeoff coefficient $\alpha=10e-6$.
  
  \subsection{Implementation of searching on Unrestricted Space}
  We use Spatial Attack, SPSA Attack and 17 Corruption  Attacks (overall 19 base attacks) for unrestricted attack policy search. For avoiding the unfairness that using unseen attacks, all the base attackers are from Unrestricted Adversarial Examples Challenge\footnote{https://github.com/google/unrestricted-adversarial-examples}. The top2 defenses (LLR, TRADESv2) in the leaderboard adopted in this work have nearly 100\% accuracy for the all base attackers. For the unrestricted attack experiment, we did not search the $\epsilon$ and $t$. We keep the magnitude of each base attacker consistent with which in the challenge. Only choices and orders of attack operations are searched. Therefore, in this case, $\alpha$ in $\mathcal{L}$ equals to 0.
  
\section{Appendix B: Implementation and analysis of Search Strategies}

\subsection{Implementation of search strategies}
For Random Search-100, we simply random sample from the candidate pool to form the attack sequence, and randomly generate their hyper-parameters. Then we choose the best from 100 such random attack policies. 

For Bayesian Optimization, we use the implementation in Scikit-Optimize\footnote{https://github.com/scikit-optimize/scikit-optimize} Library. We choose Random Forest as the base estimator. 10 initialization policies are evaluated before approximating the scoring function with base estimator. Negative expected improvement is adopted as the function to minimize over the posterior distribution. Then the search process is conducted by calling the ask\&tell methods circularly.

Reinforcement Learning is applied for data augmentation policy search in AutoAugment. This search scheme is also applicable in our case. Similar with AutoAugment, we use a one-layer LSTM as the controller, which has 100 hidden units. After some convolutional cells, the controller outputs the softmax predictions of a possible attack policy. Then the attack policy is tested on target model and get the reward. Proximal
Policy Optimization algorithm~\cite{schulman2017proximal} is used for training the controller, with the learning rate of 0.00035. After training converges, we adopt the policy produced by controller as the final result.

We use NSGA-II genetic algorithm as our final policy search scheme in CAA. Compared to Bayesian Optimization and Reinforcement Learning, NSGA-II is training free and more stable. NSGA-II needs to update a generation of attack policies. In our implementation, we set the initialized population size $K=20$, and the maximum number of generations $G=40$. The number of offsprings in each generation is $10$. The NSGA-II implementation is based on pymoo\footnote{https://github.com/msu-coinlab/pymoo} Library~\cite{pymoo}. We use tournament selection, point crossover, integer from float mutation as our selection, crossover and mutation method in NSGA-II.

\begin{figure}[!htb]
\centering
\includegraphics[width=8.3cm]{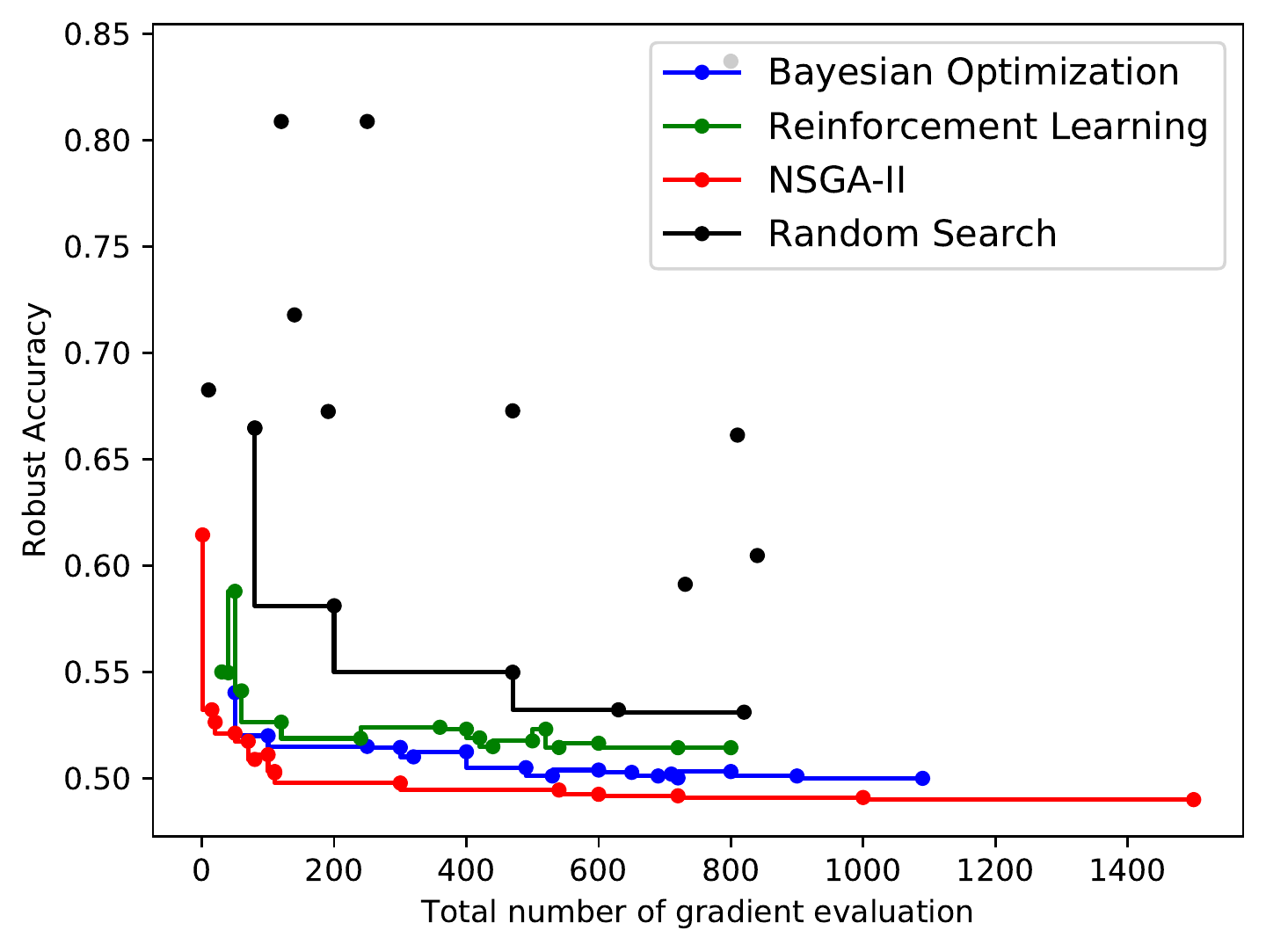}

\caption{Pareto-front of attack policies explored by NSGA-II, bayesian optimization, reinforcement learning and random search}
\label{fig:appendix1}
\end{figure}

Figure~\ref{fig:appendix1} presents the pareto-front of attack policies searched by four algorithms. The performance gap is obvious. Random search, which use 100 random policies for trials, is the farthest away to the pareto-front. As a widely used hyper-parameter optimization method, bayesian optimization algorithm outperforms the reinforcement learning. NSGA-II is the winner of these four search algorithm. It successfully find the most efficient policy with stronger attack ability under small complexity. 

\section{Appendix C: The supplementary notes of the experiments in Ablations}

\begin{table}[t] \caption{The robust accuracy (\%) tested on different model architectures and datasets. Note that CAA$_{sub}$ is searched on the proxy task that attacking ResNet-50 adversarially trained on CIFAR-10, while CAA$_{dic}$ is directly searched on the task listed in the table.}
		\label{tab:appendix1}
	\vspace{2mm}
	\centering
		%\begin{small}
		%\centering
		\small
		\begin{tabular}{l ||ccc}
			 \specialrule{0.1em}{3pt}{3pt}
			\textbf{Models} & Baseline & CAA$_{sub}$ & CAA$_{dic}$ \\
			\specialrule{0.1em}{3pt}{3pt}
		    VGG16 &
			42.41 & 42.22 & \textbf{42.13} \\
			DenseNet &
			50.27 & 49.90 & \textbf{49.90}\\
			Inception &
			48.12 & 47.67 & \textbf{47.61}\\
		\specialrule{0.1em}{3pt}{3pt}
		\textbf{Datasets} & Baseline & CAA$_{sub}$ & CAA$_{dic}$ \\
		\specialrule{0.1em}{3pt}{3pt}
					SVHN &
			56.36 & 55.91 & \textbf{55.74} \\
					MNIST &
			88.61 & 88.32 & \textbf{88.24} \\
					CIFAR-100 &
			17.84 & 17.59 & \textbf{17.54} \\
			\specialrule{0.1em}{3pt}{3pt}
	\end{tabular}\end{table}

\subsection{White-box transfer experiment of CAA}
To validate the transferability of our attack policies, we experiment CAA$_{sub}$ on diverse models and datasets (as shown in Table~\ref{tab:appendix1}). For VGG16, DenseNet and Inception, we adversarially train them on CIFAR-10. For SVHN, MNIST and CIFAR-100, ResNet-50 is used for adversarial training. We refer to \textit{robustness}\footnote{https://github.com/MadryLab/robustness} Library to implement these six tasks (the used model architecture is also provided in \textit{robustness}). AutoAttack~\cite{croce2020reliable} is used as baseline to attack these six defenses. Although CAA$_{sub}$ is transferred from other models or tasks. We show that CAA$_{sub}$ policies still have strong attack ability. The policy learned to attack ResNet50 for CIFAR-10 task leads to the improvements to attack all of the other model architectures trained on full CIFAR-10. Similarly, a policy learned on attacking CIFAR-10 task also works for other datasets that have different data and class distributions. However, compared to CAA$_{sub}$, the improvement of CAA$_{dic}$ is subtle. Particular, both CAA$_{sub}$ and CAA$_{dic}$ on DenseNet get 49.9\%. It means that policy searched on DenseNet is equal to the proxy policy. In this case, search from scratch is not necessary because it gains too negligible benefit.

\begin{table}[t] \caption{The target attack accuracy (\%) tested on different defense models. The listed defenses are built on CIFAR-10 with max $l_{\infty}=8/255$. Here CAA$_{sub}$ is not applicable in target attack, so we only compare CAA$_{dic}$ in this experiment.}
		\label{tab:appendix3}
	\vspace{2mm}
	\centering
		%\begin{small}
		%\centering
		\small
		\begin{tabular}{l ||ccc}
			 \specialrule{0.1em}{3pt}{3pt}
			\textbf{Defenses} & Baseline & CAA$_{dic}$ \\
			\specialrule{0.1em}{3pt}{3pt}
		    AdvTrain &
			10.29 & \textbf{10.45}  \\
			TRADES &
			7.39 & \textbf{7.54} \\
			AdvPT &
			6.34 & \textbf{6.59} \\
		\specialrule{0.1em}{3pt}{3pt}
	\end{tabular}\end{table}

\subsection{Target attack experiment of CAA} 
We experiment our CAA on three defense models: AdvTrain, TRADES and AdvPT. To best our knowledge, most strong attackers (\textit{e.g.}, MultiTargeted, AutoAttack) are not implemented corresponding targeted version. Therefore, we adopt PGD as the baseline in this experiment. The result is shown in Table~\ref{tab:appendix3}, our CAA$_{dic}$ also improves the success rate of the targeted attacks. It is worth noting that in the search space MT-LinfAttack is not working for targeted attack. Therefore, we delete it during search period.

\begin{table*}[t]
	\vspace{2mm}
	\centering
		%\begin{small}
		%\centering
		\small
		\begin{tabular}{l || c}
	    	\specialrule{0.1em}{3pt}{3pt}
	    	\multicolumn{2}{c}{\textbf{Attack policy searched on ImageNet adversarial training models}} \\
	    	\specialrule{0.1em}{3pt}{3pt}
	    	$S_{l_{\infty}}$ & [('MT-LinfAttack', $\epsilon$=4/255, $t$=100), ('MT-LinfAttack', $\epsilon$=4/255, $t$=50), ('PGD-LinfAttack', $\epsilon$=3/255, $t$=150)] \\ \\
	    	$S_{l_{2}}$ & [('MT-LinfAttack', $\epsilon$=3, $t$=100), ('PGD-LinfAttack', $\epsilon$=2.625, $t$=125), ('DDNAttack', $t$=1000)] \\
	    	
	    	\specialrule{0.1em}{3pt}{3pt}
	    	\multicolumn{2}{c}{\textbf{Attack policy searched in black-box transferability experiment}} \\
	    	\specialrule{0.1em}{3pt}{3pt}
			R$\rightarrow$V & [('FGSMAttack', $\epsilon$=3/255, $t$=1), ('MI-LinfAttack', $\epsilon$=8/255, $t$=125), ('CW-LinfAttack', $\epsilon$=8/255, $t$=50)] \\ \\
			R$\rightarrow$I & [('FGSMAttack', $\epsilon$=2/255, $t$=1), ('MI-LinfAttack', $\epsilon$=8/255, $t$=75), ('MI-LinfAttack', $\epsilon$=8/255, $t$=75)] \\ \\
			V$\rightarrow$R & [('FGSMAttack', $\epsilon$=2/255, $t$=1), ('IdentityAttack'), ('MI-LinfAttack', $\epsilon$=8/255, $t$=100)] \\ \\
			V$\rightarrow$I & [('FGSMAttack', $\epsilon$=3/255, $t$=1), ('CW-LinfAttack', $\epsilon$=8/255, $t$=25 ), ('MI-LinfAttack', $\epsilon$=8/255, $t$=125)] \\ \\
			I$\rightarrow$R & [('FGSMAttack', $\epsilon$=3/255, $t$=1), ('MI-LinfAttack', $\epsilon$=8/255, $t$=100 ), ('MI-LinfAttack', $\epsilon$=8/255, $t$=125)] \\ \\
			I$\rightarrow$V & [('FGSMAttack', $\epsilon$=3/255, $t$=1), ('MI-LinfAttack', $\epsilon$=8/255, $t$=100 ), ('MI-LinfAttack', $\epsilon$=8/255, $t$=100)] \\
		\specialrule{0.1em}{3pt}{3pt}
		
		\multicolumn{2}{c}{\textbf{Attack policy searched in white-box transferability experiment}} \\
		\specialrule{0.1em}{3pt}{3pt}
		VGG16 & [(’MT-LinfAttack’,$\epsilon$=8/255,$t$=50), (’MT-LinfAttack’,$\epsilon$=8/255,$t$=50), (’PGD-LinfAttack’,$\epsilon$=7/255,$t$=50)] \\ \\
			DenseNet & [(’MT-LinfAttack’,$\epsilon$=8/255,$t$=50), (’MT-LinfAttack’,$\epsilon$=8/255,$t$=25), (’CWLinfAttack’,$\epsilon$=8/255,$t$=125)] \\ \\
			Inception & [(’MT-LinfAttack’,$\epsilon$=8/255,$t$=75), (’MT-LinfAttack’,$\epsilon$=7/255,$t$=50), (’PGD-LinfAttack’,$\epsilon$=7/255,$t$=50)] \\ \\
			SVHN & [(’MT-LinfAttack’,$\epsilon$=4/255,$t$=75), (’MT-LinfAttack’,$\epsilon$=4/255,$t$=50), (’CW-LinfAttack’,$\epsilon$=4/255,$t$=100)] \\ \\
			MNIST & [ (’MT-LinfAttack’,$\epsilon$=0.3,$t$=50),(’SPSALinfAttack’,$\epsilon$=0.3,$t$=250), (’MT-LinfAttack’,$\epsilon$=0.3,$t$=50)] \\ \\
			CIFAR-100 & [(’MT-LinfAttack’,$\epsilon$=8/255,$t$=75), (’MT-LinfAttack’,$\epsilon$=8/255,$t$=25), (’MI-LinfAttack’,$\epsilon$=8/255,$t$=125)] \\
			\specialrule{0.1em}{3pt}{3pt}
		
		\multicolumn{2}{c}{\textbf{Attack policy searched for different policy length}} \\
		\specialrule{0.1em}{3pt}{3pt}
		$N=1$ & [('MT-LinfAttack', $\epsilon=8/255$, $t=100$)] \\ \\
		$N=2$ & [('MT-LinfAttack', $\epsilon=8/255$, $t=50$), ('CW-LinfAttak', $\epsilon=8/255$, $t=100$)] \\ \\
		$N=3$ & [('MT-LinfAttack', $\epsilon=8/255$, $t=50$), ('MT-LinfAttack', $\epsilon=8/255$, $t=25$), ('CW-LinfAttak', $\epsilon$=8/255, $t$=125)] \\ \\
		\multirow{2}{0.4in}{$N=5$} & [('MT-LinfAttack', $\epsilon=8/255$, $t=50$), ('MT-LinfAttack', $\epsilon=8/255$, $t=25$), ('CW-LinfAttack', $\epsilon$=6/255, $t$=100),
		\\ 
		& ('PGD-LinfAttack', $\epsilon$=2/255, $t$=25),('FGSMAttack', $\epsilon$=1/255, $t$=1)]\\ \\
		\multirow{3}{0.4in}{$N=7$} & [('MT-LinfAttack', $\epsilon=8/255$, $t=50$), ('MT-LinfAttack', $\epsilon=8/255$, $t=25$), ('CW-LinfAttack', $\epsilon$=8/255, $t$=100),\\
		& ('MT-LinfAttack', $\epsilon=7/255$, $t=25$),('CW-LinfAttack', $\epsilon=3/255$, $t=50$),('PGD-LinfAttack', $\epsilon=1/255$, $t=25$),\\
		& ('PGD-LinfAttack', $\epsilon=1/255$, $t=25$)]\\
		\specialrule{0.1em}{3pt}{3pt}
		\multicolumn{2}{c}{\textbf{Attack policy searched in target attack experiment}} \\
		\specialrule{0.1em}{3pt}{3pt}
		AdvTrain & [('MI-LinfAttack', $\epsilon=8/255$, $t=100$), ('PGD-LinfAttack', $\epsilon=7/255$, $t=75$), ('PGD-LinfAttack', $\epsilon$=7/255, $t$=75)] \\ \\
		TRADES & [('MI-LinfAttack', $\epsilon=8/255$, $t=100$), ('MI-LinfAttack', $\epsilon=8/255$, $t=125$), ('PGD-LinfAttack', $\epsilon$=7/255, $t$=75)] \\ \\
		AdvPT & [('MI-LinfAttack', $\epsilon=8/255$, $t=100$), ('PGD-LinfAttack', $\epsilon=7/255$, $t=75$), ('PGD-LinfAttack', $\epsilon$=7/255, $t$=75)] \\
		\specialrule{0.1em}{3pt}{3pt}
		
	\end{tabular}
	
	\caption{Visualization of policies founded by our CAA in each experiment. The left column shows the different settings in the corresponding experiment, and right column exhibits the best attack policy we find. For $l_{\infty}$ and $l_{2}$ attack policy searched on ImageNet, we use $l_{\infty}=4/255$ and $l_{2}=3$. For SVHN, MNIST and CIFAR-10 datasets, we search $l_{\infty}$ attack policy with max norm of $4/255$, $0.3$ and $8/255$ respectively. Since only three attackers can be shown per line, we feed a newline to represent the sequence that follows the previous line. In other cases, if not specified, we set the default max length of attack policy to 3, the type to $l_{\infty}$, and the max norm to
$8/255$.
 }
 \label{tab:appendix4}
 \end{table*}
 
\section{Appendix D: Visualization of Policies}
We visualize the policies searched by CAA in Table~\ref{tab:appendix4}.

\section{Appendix E: Visualization of Adversarial examples}
\begin{figure*}[!htb]
\centering
\includegraphics[width=17cm]{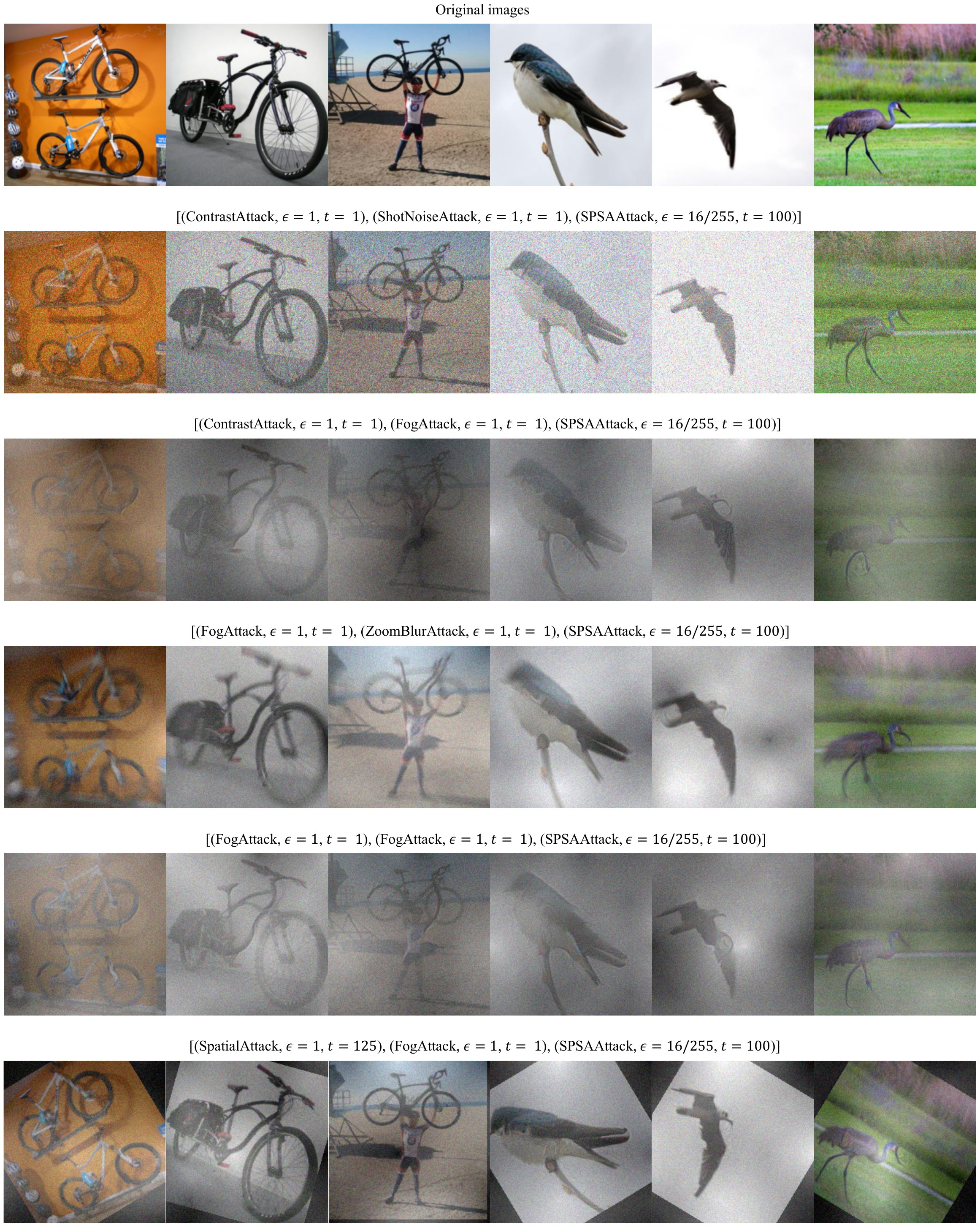}

\caption{The visualization of unrestricted adversarial examples generated by five strong attack policies searched by our CAA.}
\label{fig:appendix2}
\end{figure*}

For unrestricted attacks, an adversarial example must be guaranteed that it is unambiguous, that is, clearly decided by an ensemble of human judges. In order to demonstrate the clarity of the generated example. In Figure~\ref{fig:appendix2}, we show some adversarial examples which successfully break model. The first row represents the original images, and the next five lines each describes the adversarial examples generated by an attack policies searched by CAA. We show that without semantic confusion, the adversarial examples crafted by CAA still have strong attack ability.

\end{document}